\documentstyle[epsf,11pt,titlepage]{article}

\newcounter{nref}
\newcommand{\bbib}{%
  \renewcommand{\refname}{\large\bf References}%
  \begin{thebibliography}{99}%
  \setcounter{enumiv}{\thenref}}
\newcommand{\ebib}{%
  \end{thebibliography}%
  \setcounter{nref}{\arabic{enumiv}}}
\newcommand{\head}[3]{%
  \setcounter{nref}{0}%
  \thispagestyle{empty}%
  \section*{\LARGE\bf #1}%
  \stepcounter{section}%
  \addcontentsline{toc}{section}{#1}%
  \large\itshape%
  #2\\\vspace{0.1pt}\\%
  #3%
  \normalsize\upshape%
  \bigskip}

\begin{document}

\def\ref{\par\noindent\hangindent 15pt}


\head{Light--Element Nucleosynthesis: Big Bang and Later on}
     {\ J. L\'opez--Su\'arez $^1$ and \ R. Canal $^{1,2}$}
{$^1$ Dept. Astronomy, Univ. Barcelona, Spain\\
		       $^2$ IEEC/UB, Barcelona, Spain}

Production of the light nuclides D, $^{3}$He, $^{4}$He, and $^{7}$Li in their
currently inferred primordial abundances by standard, homogeneous big bang
nucleosynthesis (SHBBN) would imply a baryon fraction of the cosmic closure
density $\Omega_{b}$ in the range:

$$0.04\leq\Omega_{b}h^{2}_{50}\leq 0.08\eqno(1)$$

\noindent
where $h_{50}$ is the Hubble constant in units of 50 km s$^{-1}$ Mpc$^{-1}$
\cite{canal.1}, \cite{canal.2}. Therefore, for the long--time most generally
favored cosmological model with critical matter density and zero cosmological
constant ($\Omega_{M} = 1$, $\Omega_{\Lambda} = 0$), more than 90\% of the
matter in the universe should be nonbaryonic. That has led to explore
different alternatives to SHBBN (see \cite{canal.3} for a review).

Baryon inhomogeneities generated in a first--order quark--hadron phase
transition \cite{canal.4} and resulting in regions with different $n/p$
ratios has been the most thoroughly explored alternative. Agreement with the
inferred primordial abundances could only be obtained for $\Omega_{b}$ within
the range (1) again, for spherically condensed fluctuations at least
\cite{canal.5}. Recently, however, it has been shown that
$\Omega_{b}h^{2}_{50}$ might be as high as $\simeq 0.2$ in inhomogeneous
models if one assumes cylindrical shape for the inhomogeneities together with
a very high density contrast \cite{canal.6}.

Another approach has been to assume that there are unstable particles $X$,
with masses $m_{x}$ higher than a few GeV and lifetimes $\tau_{x}$ longer
than the standard thermonuclear nucleosynthesis epoch \cite{canal.7}.
Gravitinos produced during reheating at the end of inflation might be an
example of such particles. Their decay would give rise to both
electromagnetic and hadron cascades, and the resulting high--energy photons
would mainly photodisintegrate a fraction of the preexisting He whilst the
high--energy hadrons would produce light nuclides via spallation--like
reactions. A caveat of this model is that it predicts $^{6}Li/^{7}Li\gg 1$
whereas observations show that $^{6}Li/^{7}Li\leq 0.1$.

Concerning SHBBN, recent determinations of D abundances in high--redshift QSO
absorbers, when confronted with the currently inferred primordial $^{4}$He
abundance, might be in conflict with the predictions for $N_{\nu} = 3$
\cite{canal.8}. That suggests a temporary abandon at least of SHBBN as a
criterion to set bounds to $\Omega_{b}$. On the other hand, values of
$\Omega_{M}$ much lower than the closure density are now being derived from a
variety of sources, including high--$z$ supernova searches \cite{canal.9},
\cite{canal.10}. The questions of which fraction of $\Omega_{M}$ could be
baryonic and of the primordial nucleosynthesis bounds are thus posed in new
terms.

Here we explore a composite model: baryon inhomogeneities are first produced
at some phase transition prior to thermonuclear nucleosynthesis. The latter,
therefore, takes place in two different types of regions: neutron--rich and
neutron--poor ones. Then, when the universe has cooled down further and
thermonuclear reactions do no longer take place, X--particle decay starts and
the resulting electromagnetic and hadronic showers modify the light--nuclide
abundances in both regions.

We model the inhomogeneities in a very simple way: there are two types of
regions characterized by their density contrast $R$ and by their respective
volume fractions $f_{v}$ and $1 - f_{v}$. Their comoving length scale $(d/a)$
enters in the neutron diffusion rate and is a third parameter of the model.
The treatment is the same as in \cite{canal.11}. The X--particles, in turn,
are characterized by their half--life $\tau_{x}$, their mass $m_{x}$, the
ratio of their number density to that of photons $r\equiv n_{x}/n_{\gamma}$,
plus their mode of decay. The product $rm_{x}$ enters in the model as one of
the parameters, together with $\tau_{x}$. The last parameter is the effective
baryon ratio $r^{*}_{b}$, which takes into account the dependence of the
number of baryons produced in the decays on $m_{x}$ together with the
dependence of the light--element yields on the kinetic energies of the
primary shower baryons. A more detailed account of the model can be found in
\cite{canal.12} and \cite{canal.13}.

We have explored the parameter space of our model and found good agreement
with currently inferred primordial abundances for:

\smallskip

\ref
$a)$ Density contrasts $500\leq R\leq 5000$.

\ref
$b)$ Volume fractions $0.144\leq f_{v}\leq 0.192$.

\ref
$c)$ Comoving length scales $(d/a)\simeq 10^{7.5}\ cm\ Mev$ (little neutron
diffusion).

\ref
$d)$ Small abundances of the X--particles: $1.5\times 10^{-12} GeV \leq
rm_{x}\leq 1.5\times 10^{-11} GeV$.

\ref
$e)$ Half--lives of the X--particles: $6.19\times 10^{5} s\leq \tau_{x}\leq
7.43\times 10^{5} s$.

\ref
$f)$ Moderate numbers and energies of the shower baryons: $1.5\times
10^{-12}\leq r^{*}_{B}\leq 1.5\times 10^{-11}$.

\ref
$g)$ Baryon density parameter: $18\leq \eta_{10}\leq 22$.

\smallskip

Those results are illustrated in Figure 1, where we show the predicted
primordial abundances of the light nuclides as a function of $\tau_{x}$, for
$\eta_{10} = 18$, and fixed values of the other parameters taken within the
intervals a)--d) and f). The $\eta_{10}$ range translates into:

$$0.25\leq \Omega_{b}h^{2}_{50}\leq 0.35\eqno(2)$$

\noindent
in sharp contrast with (1).

\begin{figure}[ht]
\centerline{\epsfxsize=0.6\textwidth\epsffile{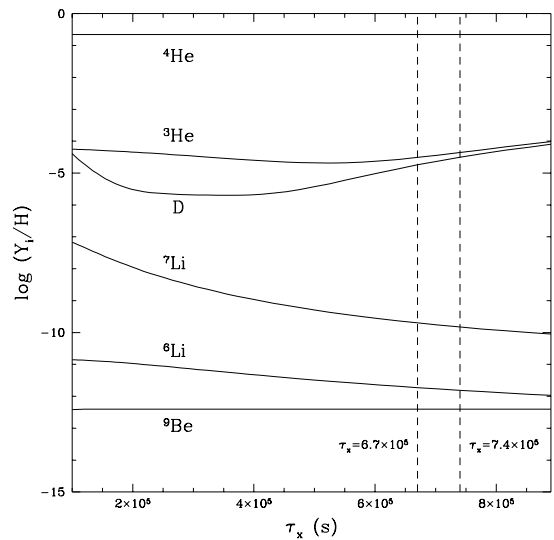}}
  \caption{Primordial abundances of the light nuclides as a function of
$\tau_{x}$, the half--life of the X--particles, for fixed values of the
other parameters}
\end{figure}

As we also see in the Figure, a testable prediction of the model is the
production of a Be abundance $(^{9}Be/H)_{p}\sim 10^{-13}$. The predicted B
abundance is much smaller. Production of Be and B is a typical feature of
inhomogeneous models. Data on Be and B abundances in halo stars now extend
down to metallicites $[Fe/H]\sim -3.0$ and they show a nearly constant B/Be
ratio $\sim 10$ \cite{canal.14} while the smallest Be abundances measured are
already of the order of our model prediction. Agreement with the observations
would thus require a reversal in the B/Be ratio at still lower metallicities.
On the other hand, the apparently primary behaviour of the Be and B
abundances in the Galactic halo, together with the Li abundances there, is a
still usolved puzzle \cite{canal.15}.

The model presented here is an example of how comparatively minor deviations
from SHBBN might very significantly broaden the range of $\Omega_{b}$
compatible with the primordial abundances inferred from observations. Planned
improvements of the model are a more realistic treatment of the
inhomogeneities, and also consideration of shorter $\tau_{x}$ for which
X--particle decays would occur simultaneously with the thermonuclear
reactions.


\end{document}